\documentstyle[preprint,aps]{revtex}
\tightenlines

\begin{document}
\newcommand{\beq}{\begin{equation}}
\newcommand{\eeq}{\end{equation}}
\newcommand{\beqa}{\begin{eqnarray}}
\newcommand{\eeqa}{\end{eqnarray}}
\newcommand{\sr}{\sqrt}
\newcommand{\fr}{\frac}
\newcommand{\mn}{\mu \nu}
\newcommand{\G}{\Gamma}

\draft
\preprint{ INJE-TP-01-04, hep-th/0104159}
\title{ Holographic principle in the BDL brane cosmology }
\author{    N. J. Kim\footnote{E-mail address:
dtpnjk@ijnc.inje.ac.kr},H. W. Lee\footnote{E-mail address:
hwlee@physics.inje.ac.kr}, and Y.S. Myung\footnote{E-mail address:
ysmyung@physics.inje.ac.kr} }
\address{
Department of Physics, Graduate School, Inje University,
Kimhae 621-749, Korea}
\author{ Gungwon Kang\footnote{gwkang@post.kek.jp}}
\address{KEK Theory Group,
        High Energy Accelerator Research Organization,
        1-1 Oho, Tsukuba,
        Ibaraki 305-0801,
        Japan}

\maketitle

\begin{abstract}
We study the holographic principle in the brane cosmology.
Especially we describe how to accommodate the 5D anti de Sitter
Schwarzschild (AdSS$_5$) black hole in the Binetruy-Deffayet-Langlois
(BDL) approach of brane cosmology.
It is easy to make a connection between  a mass $M$ of the AdSS$_5$ black hole
and a conformal field theory (CFT)-radiation dominated universe on the brane in the moving
domain wall approach. But this is not established in the BDL
approach. In this case  we use two parameters $C_1, C_2$
in the Friedmann equation. These arise from  integration and
are really related to the choice of initial bulk matter.
If one chooses a  bulk energy density $\rho_B$ to
 account for a mass $M$ of the   AdSS$_5$ black hole and the static
fifth dimension, a CFT-radiation term with $\rho_{CFT} \sim M/a^{4}$
comes out from the bulk matter without introducing a localized
matter distribution on the brane.
This means that the holographic principle  can be established in
the BDL brane cosmology.
\end{abstract}
\newpage

Recently there has been much interest in the phenomenon of
localization of gravity proposed by Randall and Sundrum
(RS)~\cite{RS1,RS2}.
RS assumed
a single positive tension 3-brane and a negative bulk cosmological
constant in the five dimensional (5D) spacetime~\cite{RS2}.
 They have obtained a four dimensional (4D) localized gravity
by fine-tuning the tension of the brane to the cosmological constant.
On the other hand, several authors have studied its cosmological
implications.
The brane cosmology   contains  some important
deviation from the Friedmann-Robertson-Walker (FRW) cosmology.
One  approach is first to assume the 5D dynamic metric (that is,
BDL-metric\cite{BDL1,BDL2,KK}) which is manifestly $Z_2$-symmetric\footnote{We call
this two-sided brane world in comparison with one-sided brane world\cite{DL}.}.
Then one solves the  Einstein equation with a localized stress-energy tensor
to find the behavior of the scale factors.
 We call this   the BDL approach.

The other  approach starts with
a static configuration which is two-sided
AdSS$_5$ spaces joined by the domain wall.
 In this case  the
embedding into the MDW\footnote{Here we use the term ``moving domain wall"
loosely to refer to any 3-brane moving in 5 dimensions.} is possible by choosing a
normal vector  $n_M$ and a tangent vector $u_M$\cite{KRA}.
The domain wall separating two bulk spaces is taken to be
located at $r=a(\tau)$, where $a(\tau)$ will  be found by solving
the Israel junction condition\cite{ISR}. In this picture an observer on the MDW  will
interpret his motion through the static bulk background as
cosmological expansion or contraction. It was shown that two
approaches can describe the same spacetime evolution \cite{MSM,DD}.

More recently, the AdS/CFT correspondence has been studied in the context of MDW approach.
This is one way of   realizing   the holographic principle.
This principle was investigated in the brane cosmology with a CFT
within one-sided AdSS$_5$ black hole spacetime\cite{VER,BFV}.
It shows that the brane  carries with the bulk
information of the AdSS$_5$ black hole.
Here the brane (MDW) is considered as the  edge of an single patch of  AdSS$_5$ space.
  A brane
starts with (big bang) inside the small black hole ($\ell>r_+$), crosses the horizon, and
expands until it reaches  maximum size. And then the brane
contracts, it falls the black hole again and finally disappears (big crunch). An
observer in  AdSS$_5$-space finds two interesting moments (see two points on
the Penrose diagram in\cite{MSM}) when the brane crosses the past (future) event
horizon. Authors in ref.\cite{SV} showed that at these times the
Friedmann equation controlling the cosmological expansion (contraction)
coincides with the  Cardy-Verlinde formula which is  the entropy-energy relation for the
 4D CFT on the brane. This is a CFT-Cosmology relation.

In this paper, we will show that the above holographic principle
can be also realized within the BDL approach.
It is easy to make  connection between  a mass $M$ of the  AdSS$_5$ black hole
and a CFT-radiation dominated universe on the brane in the moving
domain wall approach\cite{VER,BFV}.
But as far as we know, this is not yet established in the BDL
approach. In the BDL method we usually have two parameters $C_1, C_2$
in the Friedmann equation\cite{BDL2}. These result from integration and
are really related to the choice of  initial bulk matter $(\rho_B,P)$.
Hence  we will use this parameter to make a connection between the
brane and the  bulk. This is possible because the BDL metric can
lead to the AdSS$_5$ metric by a coordinate transformation
:$\{\tau,x^i,y\} \to  \{t,r,\chi,\theta,\phi\}$.
We find a relation $C_2(0)=\kappa^2a^4_0\rho_B/6$ which
turn out to be constant on the brane in order to go well with other
conditions, in addition to $C_1=0$.
Considering  both $C_2(0)\sim M$ and the static
condition $\dot b=0$ for the  fifth dimension, a CFT-radiation term with
 $\rho_{CFT} \sim M/a_0^4$
comes out from the bulk spacetime without introducing a localized
matter distribution on the brane. Hence we call the BDL approach
under certain conditions
as the fixed domain wall approach, in compared with the MDW
approach.

We start with the BDL metric for the brane cosmology\cite{BDL2}

\beq
ds^2 = -n^2 (\tau,y)d\tau^2 +a^2 (\tau,y) \gamma_{ij} d x^i d x^j +b^2
(\tau,y)d y^2,
\eeq
where $\gamma_{ij}$ is a maximally symmetric 3D space parametrized
by $k\in \{-1,0,1\}$.

Then the 5D Einstein equation takes the form
\beq
G_{AB}\equiv R_{AB}-\fr{1}{2}R g_{AB}=\kappa^2 T_{AB}
\eeq
with $\kappa^2= 8 \pi G_5^N$. The Einstein tensor is given by
\beqa
&&G_{00}= 3\Big\{{\Big(\fr{\dot{a}}{a}\Big)}^2+ \fr{\dot{a}\dot{b}}{a b}
-\fr{n^2}{b^2}\Big(\fr{a''}{a}+\Big(\fr{a'}{a}\Big)^2 -\fr{a' b'}{a b}
\Big)+ k \fr{n^2}{a^2}\Big\},\cr
&&G_{i j}=\fr{a^2}{b^2}\gamma_{ij}
\Big\{ {\Big(\fr{a'}{a}\Big)}^2 +2\fr{a' n'}{a n}-2\fr{a'b'}{a b}
-\fr{b' n'}{b n} +2\fr{a''}{a}+\fr{n''}{n}\Big\} \cr
&&+\:\:\fr{a^2}{n^2}\gamma_{ij}\Big\{ -\Big(\fr{\dot{a}}{a}\Big)^2 +2\fr{\dot{a} \dot{n}}{a n}
-2\fr{\dot{a}\dot{b}}{a b}+\fr{\dot{b}\dot{n}}{b n}
-2\fr{\ddot{a}}{a}+\fr{\ddot{b}}{b}
\Big\}-k\gamma_{ij},\cr
&&G_{05}=3\Big(\fr{\dot{a} n'}{a n}+\fr{a' \dot{b}}{a
b}-\fr{\dot{a}'}{a}\Big),\cr
&&G_{55}=3\Big\{{\Big(\fr{a'}{a}\Big)}^2+ \fr{a' n'}{a n}
-\fr{b^2}{n^2}\Big(\fr{\ddot{a}}{a}+\Big(\fr{\dot{a}}{a}\Big)^2
-\fr{\dot{a}\dot{n}}{a n}\Big)-k \fr{n^2}{a^2}
\Big\},
\eeqa
where dot(prime) denote differentiations with respect to
$\tau(y)$.
And  the  matter source is given by the bulk and brane parts
\beq
{T^A}_B ={\check{T}^A}\, _{B} |_{\mbox{\scriptsize{bulk}}} +{\tilde{T}^A}\,_{B}
|_{\mbox{\scriptsize{brane}}}
\eeq
with the bulk stress-energy tensor distributed in whole space
\beq
{\check{T}^A}\, _{B} |_{\mbox{\scriptsize{bulk}}} =\left[
-(\rho_B +\Lambda), -\Lambda +P, -\Lambda +P, -\Lambda +P, -\Lambda +P_{\scriptsize{T}}\right]
\eeq
and the localized matter on the  brane at $y=0$
\beq
{\tilde{T}^A}\, _{B} |_{\mbox{\scriptsize{brane}}} =\fr{\delta (y)}{b}\left[
-(\sigma +\rho_b), -\sigma +p_b, -\sigma +p_b
, -\sigma +p_b,0 \right].
\eeq
Here we split the bulk tensor into the cosmological constant $\Lambda$ and the
 bulk  matter ($\rho_B,P,P_T$) for convenience.
The brane tension $\sigma$ is introduced for the fine-tuning.
Also we assume that all of matters are independent
of the fifth coordinate $y$.
Further we include
the different bulk pressure $P_T$ along the fifth direction for general discussion.
The assumption  of ${\tilde{T}^0}\, _{5}=0$ which means that there
is no flow of matter along the fifth direction implies that
$G_{05}=0$.
Let us introduce $F$ to solve the Einstein equation as
\beq
F\equiv\fr{(a a')^2}{b^2}-\fr{(a \dot{a})^2}{n^2}-ka^2.
\eeq
Then  we note that two components of Einstein equation take the
forms
\beqa
&&G_{00}=-\fr{3}{2}\fr{n^2}{a^3 a'}F'=\kappa^2 T_{00},\\
&&G_{55}=\fr{3}{2}\fr{b^2}{a^3 \dot{a}}\dot{F}=\kappa^2 T_{55}
\eeqa
which imply
\beqa
&&F'=-\fr{2}{3}\fr{a^3 a'}{n^2}\kappa^2 T_{00}= -\fr{2}{3}\kappa^2
a^3 a'\big( \Lambda +\rho_B +\fr{1}{b}(\sigma
+\rho_{b})\delta(y)\big),\label{FPR}\\
&&\dot{F}=\fr{2}{3}\fr{a^3 \dot{a}}{b^2}\kappa^2 T_{55}= -\fr{2}{3}\kappa^2
a^3 \dot{a}\big( \Lambda -P_{T}\big).
\label{FDO}
\eeqa

In the bulk of
 which is  everywhere except $y\not=0$, Eqs.(\ref{FPR}) and
 (\ref{FDO}) lead to
\beqa
&&F'= -\fr{2}{3}\kappa^2 a^3 a' (\Lambda +\rho_B),
\label{FPR1} \\
&&\dot{F}= -\fr{2}{3}\kappa^2 a^3 \dot{a}(\Lambda -P_{T})\label{FDO1}.
\eeqa

Integrating Eq.(\ref{FPR1}) over $y$ and
Eq.(\ref{FDO1}) over $\tau$ give us two integration constants $C_1(\tau)$
and $C_2(y)$\footnote{Compared with\cite{BDL2,DEFF}, $C_1$ is related to a free
parameter ${\cal C}$ which  may carry with the mass $M$ of AdSS$_5$ black hole,
whereas $C_2$ is not considered in the BDL case. In this work we will determine these
by both the Einstein equation and the bulk  conservation law.}

\beqa
&&F=-\fr{1}{6}\kappa^2 a^4(\Lambda+\rho_B)-C_{1}(\tau)\label{FR1},\\
&&F=-\fr{1}{6}\kappa^2 a^4(\Lambda-P_T )-\fr{1}{4}\int a^4\dot{P_T} d
\tau-C_{2}(y)\label{FR2}
\eeqa
which combine to find out the relation between $C_1(\tau)$ and $C_2(y)$

\beq
C_{2}(y)=\fr{1}{6}\kappa^2 a^4(\rho_B + P_T )-\fr{1}{4}\int a^4\dot{P_T} d
\tau +C_{1}(\tau).
\eeq
This implies that the first term of the  r.h.s is independent of $\tau$ and
 $C_1$ is just constant because the l.h.s. is only  a
function of $y$. Plugging the definition of $F$ into
Eqs.(\ref{FR1}) and (\ref{FR2}), one has two apparently different
equations
\beqa
&&\Big( \fr{\dot{a}}{n a}\Big)^2 =-\fr{k}{a^2}+ \Big( \fr{a'}{b a}\Big)^2
+ \fr{1}{6}\kappa^2 (\Lambda+\rho_B)+\fr{C_{1}(\tau)}{a^4},\label{EQ0}\\
&&\Big( \fr{\dot{a}}{n a}\Big)^2 = -\fr{k}{a^2}+\Big( \fr{a'}{b a}\Big)^2
+ \fr{1}{6}\kappa^2 (\Lambda-P_T )+\fr{1}{4}\int a^4\dot{P_T}
d\tau
+\fr{C_{2}(y)}{a^4}\label{EQ1}.
\eeqa

First we consider  Eq.(\ref{EQ1}) because Eq.(\ref{EQ0}) will give
us the same equation as in Eq.(\ref{EQ1}).
redundant.
We are in a position to account the localized distribution on
the brane at $y=0$ using the Israel junction condition. From (0,0)-and
$(i,j)$-components of the Einstein equation, we obtain

\beqa
&& \fr{[a']}{a_0 b_0}=-\fr{1}{3}\kappa^2(\sigma +\rho_b ), \\
&& \fr{[n']}{n_0 b_0}=\fr{1}{3}\kappa^2(-\sigma +2\rho_b +3 p_b )
\eeqa
with $[a']\equiv a'(0^{+})-a'(0^{-})$. On the brane we have the
corresponding conservation law
\beq
\dot{\rho_b }+3(\rho_b +p_b )\fr{\dot{a_0}}{a_0}=0,
\eeq
where the subscript "0" means the location of the brane at $y=0$.
Eq.({\ref{EQ1})
 should also be satisfied as one approaches the
brane, i.e., $y\to 0^{+}.$

\beq
\Big( \fr{\dot{a_0}}{n_0 a_0}\Big)^2 =-\fr{k}{a_0^2}+ \Big( \fr{a'(0^{+})}{b_0 a_0}\Big)^2
+ \fr{1}{6}\kappa^2 (\Lambda-P_T )+\fr{1}{4}\int a_0^4\dot{P_T} d t
+\fr{C_{2}(0)}{a_0^4}
\eeq
Assuming the $Z_2$-symmetry of $a(-y)=a(y),n(-y)=n(y)$, we have
\beqa
&&\fr{a'(0^{+})}{b_0 a_0}= -\fr{1}{6}\kappa^2 (\sigma+ \rho_b ),\\
&&\fr{n'(0^{+})}{b_0 n_0}= \fr{1}{6}\kappa^2 (-\sigma+2\rho_b + 3p_b
).
\eeqa
Redefining the time as $n_0 (\tau,y=0)d \tau\equiv d\tau$ (i.e., $n_0 \equiv
1$), one has the Friedmann-like equation on the brane
\beq
H^2 \equiv \Big(\fr{\dot{a_0}}{a_0}\Big)^2 =-\fr{k}{a_0^2}+\fr{\kappa^4
\sigma^2}{36}+\fr{\kappa^2}{6}(\Lambda -P_T ) + \fr{1}{4}\int a_0^4\dot{P_T}d\tau
+\fr{\kappa^4 \sigma}{18}\rho_b + \fr{\kappa^4}{36}\rho_b\,^2 +
\fr{C_{2}(0)}{a_0\,^4}.
\eeq
For the fine-tuning, if one chooses
$\Lambda= -\fr{1}{6}\kappa^2 \sigma^2=-\fr{6}{\ell^2\kappa^2
}$ with $\sigma=\fr{6}{\ell\kappa^2}$
then the above equation leads to
\beq
H^2 =-\fr{k}{a_0^2}-\fr{\kappa^2}{6}P_T
+ \fr{1}{4}\int a_0^4\dot{P_T} d\tau
+\fr{\kappa^2}{3\ell}\rho_b + \fr{\kappa^4}{36}\rho_b\,^2 +
\fr{C_{2}(0)}{a_0\,^4}.
\label{EQH}
\eeq
Without the localized matter $(\rho_b=0)$, Eq.(\ref{EQH}) reduces to
\beq
H^2 =-\fr{k}{a_0^2}-\fr{\kappa^2}{6}P_T
+ \fr{1}{4}\int a^4_0\dot{P_T} d\tau
+ \fr{C_{2}(0)}{a_0\,^4}.
\label{EQH1}
\eeq
Now we wish to use  Eqs.(\ref{FPR1}) and
(\ref{FDO1}) to obtain further information.
Considering $\dot{(F')}=(\dot{F})'$ and $G_{05}=0$ gives
\beq
\dot{\rho}_B a' +\dot{a}\Big( 3\fr{a'}{a}+\fr{n'}{n}\Big)\rho_B
 +a' \fr{\dot{b}}{b}(\rho +P_{T})=0.
 \label{FFR}
\eeq
Also from (0,0)-and $(i,j)$-components of the  Einstein
equation, one finds
\beq
\rho_B \fr{n'}{n}=3P\fr{a'}{a}.
\eeq
Using the above  relation, Eq.(\ref{FFR}) leads to
\beq
\dot{\rho}_B + 3\Big( \rho_B+P +P_{T}+\fr{P}{\rho_B}P_T
\Big)\fr{\dot{a}}{a}+(\rho_B+P_T )\fr{\dot b}{b}=0.
\label{FFR!}
\eeq
This may be considered a variant of  the  bulk conservation law $\nabla_A \check T^A\,_0
=0$,

\beq
\dot{\rho}_B + 3( \rho_B+P)\fr{\dot{a}}{a}+(\rho_B+P_T )\fr{\dot b}{b}=0
\label{COL}
\eeq
which can be also derived from  the
Bianchi identity of $\nabla_A G^A\,_0=0$.
If one compares  Eq.(\ref{FFR!}) with Eq.(\ref{COL}),
Eq.(\ref{FFR!}) means the violation of the
conservation law due to the introduction of  non-zero
$P_T$. Actually as it stands, Eq.(\ref{FFR!})  can provide us another
checkpoint
to
derive a consistent brane cosmology. Here we have to choose $P_T$ in a way
 that Eq.(\ref{FFR!}) is
 consistent with  the conservation law Eq.(\ref{COL}).

 Let us discuss some interesting cases:
i) In the case of $P_T=P$,  we find an obvious violation of
conservation law.
ii) If $P_T=0$, Eq.(\ref{FFR!}) leads to the known conservation law
for the bulk matter in the brane world scenario\footnote{Here we
split the bulk tensor into the cosmological constant $\Lambda$ and the
 bulk fluid matter ($\rho_B,P,P_T$). But in ref.\cite{BDL2},
they did not introduce such a bulk  matter.}\cite{BDL2}.
In this case we  can get the brane expansion as well as the bulk
expansion.
iii) $\rho_B=-P$ leads to  the case that the brane is vacuum whereas the expansion
of  extra dimension is governed by $\rho_B \sim b^{-(1+\tilde w)}$
according to $\dot{\rho}_B + (\rho_B+P_T )\fr{\dot b}{b}=0$
with $P_T=\tilde w \rho_B$.
iv) If $P_T=-\rho_B$, we have the vacuum
state for bulk with $\dot \rho_B=0$.

For our purpose, it is desirable to choose $P_T=0$. This means that
the bulk matter does not flow into the fifth direction. This situation is similar to the
case
that ${\tilde{T}^0}\, _{5}=0$  implies
that there is no flow of a localized matter along the fifth direction.
But we remind the reader that our situation  is still under $\Lambda
\not=0$ and $\rho_B\not=0$ in the bulk direction.
Hence $P_T=0$ implies the matter-dominated universe in the bulk
direction.
In this case, comparing Eq.(\ref{EQ1}) with Eq.(\ref{EQ0})
leads to  $C_2(0)=\kappa^2a_0^4\rho_B/6+C_1$, which means that the r.h.s. is
independent of $\tau$. Let us choose  $C_1,C_2$ to obtain an
appropriate brane cosmology (brane expansion or contraction).
This means that we use two parameters $C_1,C_2$
to embed the bulk matter into the brane cosmology.
Here $C_1=0$ for our purpose.
From the fact that $C_2(0)=\kappa^2a_0^4\rho_B/6$ is constant w.r.t. $\tau$,
 we get immediately an important relation of
  $\rho_B \sim a_0^{-4}$ on the brane.

On the other hand,
using the  bulk conservation law Eq.(\ref{COL}) with $P_T=0$ , we
 can arrive at
the same result.  That is,
 $\rho_B \sim a_0^{-4}$ is recovered if $\dot b=0$ and $P=\rho_B/3$.
This means that the result of MDW approach can be recovered from the BDL method if
 the brane is  an only  expanding subspace in the  bulk spacetime.
 As a result Eq.(\ref{EQH1}) leads to
\beq
H^2  =-\fr{k}{a_0^2}
+ \fr{\kappa^2}{6}\rho_B
\label{EQH2}
\eeq
on the brane. Also it is important to check
  that from the  equation (\ref{EQ0}) with $C_1=0$, we recover the same
equation as Eq.(\ref{EQH2}).

Finally let us introduce the CFT-radiation dominated universe for one-sided brane
relation\footnote{The relevant difference between one-sided and two-sided scenarios
is to define the 4D Newtonian constant\cite{DL}: $8 \pi G_4^N=\fr{2\kappa^2}{\ell}$ for one-sided
, $8 \pi G_4^N=\fr{\kappa^2}{\ell}$ for two-sided.} with
$k=1$\cite{SV}

\beq
H^2  =-\fr{1}{a_0^2}
+ \fr{8\pi G_4^N}{3}
\rho_{CFT},~~\rho_{CFT}=\fr{E}{V}=\fr{M\ell}{a^4_0V(S^3)},~~8\pi
G_4^N=\fr{2\kappa^2}{\ell}
\label{EQH3}
\eeq
where $M$ is the bulk information (mass of  the AdSS$_5$ black hole on each side, while $E$ is
the boundary information (its energy on the brane). $V(S^3)$ is the volume of the unit  $S^3$.
Comparing Eq.(\ref{EQH2}) with Eq.(\ref{EQH3}),
one finds that if $\rho_B=2\rho_{CFT}/\ell= 2M/(a_0^4V(S^3))$, the
CFT-radiation universe recovers from the BDL approach.
In this case  we have $C_2(0)= \kappa^2 \rho_B a_0^4/6= \kappa^2M/(3 V(S^3))$.
This implies that $C_2(0) \sim M$. If $C_2=0$, the bulk spacetime
is the exact AdS$_5$ spacetime with $\Lambda$.
$C_2\not=0$ is related to the initial bulk matter distribution : the 5D non-vanishing
Weyl tenor. In the exact AdS$_5$ space, we find the vanishing Weyl
tensor, whereas we get the non-vanishing one for the AdSS$_5$ black
hole spacetime.

Let us discuss what conclusions can follow from Eq.(\ref{EQH3}).
This implies that Friedmann equation
knows about  thermodynamics of the radiation-matter CFT on the brane.
Initially the Friedmann equation has nothing to do with the CFT.
But using both the AdS/CFT correspondence and the various entropy bounds, we make a connection
between the above Friedmann equation and the entropy($S$)-energy($E$)
 relation for the CFT\cite{VER,SV}.
Explicitly we arrive from Eq.(\ref{EQH3}) exactly at the Cardy-Verlinde formula
: $ S=(2\pi a_0/3)\sqrt{E_c(2E-E_c)}$
 with the Casimir energy $E_c$, when the brane crosses the horizon of the balck
 hole.

In conclusion we establish the holographic principle  using the BDL
approach. We get the Friedmann equation which describes
the CFT radiation-dominated  universe on the
brane from the initial bulk consideration.
 Especially we use two  parameters $C_1$ and $C_2$ in the BDL approach to
embed the initial bulk configuration of the AdSS$_5$ black
hole into the brane. With $\rho_B\sim M/a_0^4$, the bulk matter satisfies
$P=\rho_B/3$ on the brane .
This  means  that the black hole with mass $M$ in the BDL method provides the CFT radiation
matter on the brane as in the case of  the moving domain wall approach. The
BDL (MDW) approaches use the Gaussian
normal (Schwarzschild) coordinates. Mukhoyama et al. showed that
although there exist apparent different configurations,
all of two-types solutions represent the same spacetime described with
different coordinate systems\cite{MSM}. Here our calculation also
shows that two approaches give us the same result. In this sense
our BDL approach can be considered as the  domain wall
approach fixed at $y=0$, in compared with the MDW approach.

\section*{Acknowledgments}
We thank R.G. Cai for helpful discussions.
This work was supported in part by the Brain Korea 21 Program, Ministry of
Education, Project No. D-0025 and KOSEF, Project No. 2000-1-11200-001-3.


\begin{thebibliography}{99}
\bibitem{RS1} L. Randall and R. Sundrum, Phys. Rev.Lett. 83 (1999) 3370
[hep-ph/9905221].

\bibitem{RS2} L. Randall and R. Sundrum, Phys. Rev.Lett. 83 (1999)
4690 [hep-th/9906064] .

\bibitem{BDL1} P. Binetruy, C. Deffayet and D. Langlois,
Nucl. Phys. B 565 (2000) 269 [hep-th/9905012].

\bibitem{BDL2} P. Binetruy, C. Deffayet, U. Ellwanger and D. Langlois, Phys. Lett. B 477
(2000) 285 [hep-th/9910219].
\bibitem{DEFF} C. Deffayet, hep-th/0010186;T. Shiromizu and D. Ida,
hep-th/0102035.


\bibitem{KK} H. B. Kim and H. D. Kim, Phys. Rev.  D61
(2000 )064003[hep-ph/9909053].
\bibitem{DL} M. J. Duff and T. Liu, Phys. Rev. Lett.  85
(2000) 2052.

\bibitem{KRA} P. Kraus, JHEP 9912 (1999) 011 [hep-th/9910149];
 D.  Ida, JHEP 0009 (2000) 014;
   H. Collins and B. Holdom, hep-ph/0003173;
  H. Stoica, S. H. H.  Tye and I. Wasserman, hep-th/0004126;
  C. Barcelo and M. Visser, hep-th/0004056;
   A. C. Davis, I. Vernon, S. C. Davis and W. B. Perkins,
  hep-ph/0008132;
 N. J. Kim, H. W. Lee, and Y. S. Myung,
hep-th/0101091.

\bibitem{ISR} W. Israel, Nuovo Cim. B44 (1966) 1; {\it ibid}.
B48 (1967) 463.

\bibitem{MSM}  S. Mukhoyama, T. Shiromizu and K. Maeda, hep-th/9912287.
\bibitem{DD} N. Deruelle and T. Dolezel, gr-qc/0004021.

\bibitem{VER} E. Verlinde, hep-th/0008140.


\bibitem{BFV} R. Brustein, S. Foffa
and G. Veneziano, hep-th/0101083; D. Klemm, A. Petkou and G.
Siopsis, hep-th/0101076; B. Wang, E. Abdalla and R. K. Su,
hep-th/0101073; S. Nojiri and S. Odintsov, hep-th/0011115; F.L.
Lin, hep-th/0010127; D. Kutasov and F. Larsen, hep-th/0009244;
R. G. Cai, hep-th/0102113; Y.S. Myung, hep-th/0102184;
D. Birmingham and S. Mokhtari, hep-th/0103108;
S. Nojiri and S. Odintsov, hep-th/0103078;
L. Anchordoqui, C. Nunez, K. Olsen, hep-th/0007064 ;
A. K. Biswas and S. Mukherji, hep-th/0102138;
Y.S. Myung, hep-th/0103241.

\bibitem{SV} I. Savonije and E. Verlinde, hep-th/0102042.










\end{thebibliography}
\end{document}